\documentclass[10pt,aps,prl,superscriptaddress,nofootinbib,nobibnotes,longbibliography,floatfix,twocolumn]{revtex4-2}

\usepackage{bm}
\usepackage{mathtools,
amsmath,
amssymb,
amsfonts,
mathrsfs,
chngcntr,
multirow}

\let\cc\corresponds
\let\corresponds\relax
\usepackage{mathabx}
\let\corresponds\cc

\usepackage[utf8]{inputenc}
\usepackage[T1]{fontenc}
\usepackage[dvipsnames]{xcolor}
\usepackage[unicode]{hyperref}
\hypersetup{colorlinks=true, citecolor=MidnightBlue,
            linkcolor=MidnightBlue, urlcolor=MidnightBlue, linktocpage=true}
\usepackage[normalem]{ulem}
\usepackage{orcidlink}
\usepackage{afterpage}
\usepackage{float}
\usepackage{placeins}
\usepackage{flafter}
\usepackage{balance}

\newcommand{\orcid}[1]{\href{https://orcid.org/#1}{\includegraphics[width=10pt]{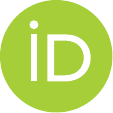}}}

\begin{document}

\title{Bound States of the Schwarzschild Black Hole}

\author{Sebastian H. V\"olkel \orcid{0000-0002-9432-7690}}
\email{sebastian.voelkel@aei.mpg.de}
\affiliation{Max Planck Institute for Gravitational Physics (Albert Einstein Institute),
D-14476 Potsdam, Germany}

\date{\today}

\begin{abstract}
Understanding the physical significance and spectral stability of black hole quasinormal modes is fundamental to high-precision spectroscopy with future gravitational wave detectors. 
Inspired by Mashhoon's idea of relating quasinormal modes of black holes with their equivalent bound states in an inverted potential, we investigate, for the first time, energy levels and eigenfunctions of the Schwarzschild black hole quantitatively. 
While quasinormal modes describe the characteristic damped oscillations of a black hole, the bound states of the inverted potential are qualitatively more similar to those of the hydrogen atom. 
Although the physical interpretation of these states may initially be of more academic interest, it furthers our understanding of open problems related to quasinormal modes in a similar spirit to Maggiore's interpretation of the Schwarzschild quasinormal mode spectrum. 
One surprising insight from the explicit calculation of bound states is that eigenfunctions corresponding to quasinormal mode overtones become rapidly delocalized and extremely loosely bound. 
This observation raises immediate questions about the common interpretation of quasinormal modes as excitations of the lightring region. 
Closely related, as a second application, we also explore the spectral stability of bound states and demonstrate that they can provide complementary insights into the quasinormal mode spectrum. 
\end{abstract}

\maketitle

\section{Introduction}\label{intro}

Black hole quasinormal modes (QNMs) describe the characteristic damped oscillations of our Universe's most extreme gravitational objects. 
They have attracted the interest of different research communities, for instance, gravitational wave astronomy by measuring the ringdown of astrophysical binary black hole mergers~\cite{Dreyer:2003bv,LIGOScientific:2016aoc,LIGOScientific:2016lio}, mathematical relativity in terms of exploring black hole stability~\cite{Dafermos:2016uzj}, and fundamental physics for testing general relativity~\cite{Berti:2018vdi,Cardoso:2019rvt,Franchini:2023eda}, as well as, searching for quantum mechanical properties of black holes~\cite{Hod:1998vk}. 
The vast amount and diversity of literature on these topics and beyond demonstrate a comprehensive interest of the community on the one hand and the existence of open questions on the other~\cite{Kokkotas:1999bd,Nollert:1999ji,Berti:2009kk,Konoplya:2011qq}. 

There are fundamental properties of the black hole QNM spectrum that are not easy to understand when 
comparing them to the spectral properties of basic mechanical systems or even the most compact stars. 
In the simplest case of the Schwarzschild black hole, its metric perturbations can be brought in the form of a master equation for a radial function $\psi(r)$ that is equivalent to solving the time-independent Schr\"odinger equation
\begin{align}\label{schrodinger}
\frac{\text{d}^2}{\text{d}x^2}\psi (x) + \left[\omega_n^2 - V_{\ell}(r) \right] \psi(x) = 0\,,
\end{align}
with a family of potential barriers $V_{\ell}(r)$ labeled by a separation constant $\ell(\ell+1)$ from the angular problem. 
Note the use of the tortoise coordinate $x(r) = r + 2\,M \ln\left(r/2\,M-1\right)$, where $M$ is the mass of the black hole.  
Depending on the parity of the metric perturbations, the potentials are either given by the Regge-Wheeler (axial) or Zerilli (polar) potentials~\cite{Regge:1957td,Zerilli:1970se}; here for the Regge-Wheeler case studied later
\begin{align}\label{potential}
V_{\ell}(r) = \left(1- \frac{2M}{r} \right) \left(\frac{\ell(\ell+1)}{r^2}-\frac{6M}{r^3} \right)\,.
\end{align}
The corresponding QNMs $\omega_n$, where $n$ labels the overtone number, are the solutions of an eigenvalue problem with purely outgoing waves at spatial infinity and purely ingoing waves at the horizon~\cite{Chandrasekhar:1975zza}. 
Despite this surprisingly simple reduction of the whole problem to a seemingly more familiar one, the solution differs from typical applications. 

QNMs of the Schwarzschild black hole are strongly damped modes whose fundamental mode, for a given $\ell$, has by definition not only the smallest imaginary part, but also the largest real part. 
Its overtones first decrease in frequency until becoming zero and then increase again until asymptotically reaching a constant value; see figure~2 in Ref.~\cite{Kokkotas:1999bd}. 
The imaginary part increases continuously and becomes uniformly spaced for large overtones~\cite{Nollert:1993zz}. 
Such behavior is different from the one of most mechanical systems, where overtone frequencies typically increase without a limit while their imaginary part grows less rapidly. 
Note that not even the QNMs of ultra-compact stars scale in such a way~\cite{doi:10.1098/rspa.1991.0104,Kokkotas:1994an}.

This somewhat puzzling structure of the QNM spectrum lead Maggiore to suggest a mapping between the black hole QNMs to those of an effective, damped oscillator~\cite{Maggiore:2007nq}. 
Its spectral properties are intuitively plausible and have, in the same work, been further used to revisit the connection of large overtones and the quantization constant in black hole area quantization~\cite{Bekenstein:1974jk,Mukhanov:1986me}.

Another mapping of the QNM spectrum was put forward earlier by Mashhoon~\cite{Mashhoon:1982im} and applied in more detail in subsequent works~\cite{BLOME1984231,Ferrari:1984ozr,Ferrari:1984zz}. 
A simple, complex coordinate transformation introduced in Eq.~\eqref{schrodinger} can be used to connect the spectral properties of QNMs with those of the bound states of the ``inverted'' potential barrier by explicitly transforming the parameters describing the potential. 
Applying these transformations requires an analytic expression for the bound states. 
Therefore, Mashhoon's approach has, until recently~\cite{Volkel:2022ewm}, been limited to cases where the bound states are known analytically, e.g., in terms of the P\"oschl-Teller potential or other exactly solvable potentials.  
This limited the method to approximate potentials and QNMs with small overtone numbers. 
In Ref.~\cite{Hatsuda:2019eoj}, the inverted potential approach was discussed in connection with the higher-order WKB method, which provides a semi-analytic result, while connections to Seiberg-Witten theory can be found in Ref.~\cite{Aminov:2020yma}. 
Formal limitations of the bound states to the QNM correspondence are discussed in Ref.~\cite{Richarte:2024bmm} 
and a related application using hyperboloidal methods can be found in Ref.~\cite{Burgess:2024hks}. 
Note that generalizations of the classical Bohr-Sommerfeld rule have also been used to compute QNMs in Ref.~\cite{Kokkotas:1991vz}. 

Because the bound-state spectra of the inverted Regge-Wheeler and Zerilli potentials are not known analytically, Mashhoon's approach has never been directly applied to these cases. 
The closest connection we are aware of is a discussion and analytical study connected to the Coulomb potential in Ref.~\cite{1996CQGra..13..233L}, which demonstrates the existence of infinitely many bound states and makes approximate predictions for the Schwarzschild QNMs. 
However, to our knowledge, the bound states and their eigenfunctions have not been reported in the literature. 
Although the inverted potential does not have an obvious physical interpretation, besides being used as a tool to compute QNMs, we argue in this work that \textit{bound states carry insightful information} and can provide a complementary understanding of the QNM spectrum and open problems around it. 
Similarly, abstract descriptions and mappings in gravitational wave research, such as the effective-one-body formalism~\cite{Buonanno:1998gg} or the scattering-to-bound mapping~\cite{Kalin:2019rwq} have proven to be powerful tools. 

As an explicit demonstration, we first report numerical results on the bound states and eigenfunction of the Regge-Wheeler potential and leave a detailed study for other fields for future work. 
We find that large eigenvalues approach zero exponentially fast, qualitatively like $E_n\sim-\exp(-K n)$, where $K$ is some constant, rather than $E_n\sim-1/n^2$ as for the hydrogen atom. 
This behavior has immediate consequences for the eigenfunctions, which are less localized and surprisingly sensitive to the effective potential far beyond the vicinity of the lightring. 

To further outline the possibility of using bound states to address open problems, we also study the impact of small perturbations in the Regge-Wheeler potential on the spectrum and the eigenfunctions. 
This application is directly related to questions of the physical significance of the QNM spectrum~\cite{Nollert:1996rf,Nollert:1998ys} and possible environmental effects~\cite{Barausse:2014tra}. 
More recently, these aspects were also studied regarding spectral stability via the pseudospectrum~\cite{Jaramillo:2020tuu}. 
For the latter, it was reported in the literature that modifications of this type destabilize the overtones. 
At the same time, the fundamental mode is more robust~\cite{Cheung:2021bol,Berti:2022xfj,Cardoso:2024mrw,Ianniccari:2024ysv,Siqueira:2025lww}, which is in agreement with the previously mentioned works. 
This behavior can be explained in the bound-state picture. 
We report that bound states below the modification's energy level are robust. 
In contrast, energy levels beyond the modification receive an overall shift, although we would not refer to them as unstable. 

In this work, we adopt units in which $G=c=1$.

\section{Methods}

Let us first review why and how QNMs are related to the spectrum of bound states. 
According to Mashhoon's approach~\cite{Mashhoon:1982im}, one can map the QNMs $\omega_n^2$ of a potential barrier with the spectrum of bound states $E_n$ of the inverted barrier. 
Applying the complex coordinate transformation $x \rightarrow -ix$ in Eq.\eqref{schrodinger}, and demanding that $V(x, P) = V(-ix, P^\prime)$, one finds that $\Omega^2(P) \equiv -E_n(P)$ of the potential well can be used to compute the QNMs via 
\begin{align}\label{bound_to_qnm}
\omega_n(P) \equiv \Omega_n(\pi^{-1}(P))\,.
\end{align}
Here, $P$ describes the parameters of the potential, and the mapping $P^\prime = \pi(P)$ depends on the specific potential. 
Instead of explicitly carrying out the mapping, we only rely on the formal aspect that QNMs and bound states can be mapped to each other, but refer to Refs.~\cite{Mashhoon:1982im,BLOME1984231,Ferrari:1984ozr,Ferrari:1984zz,Churilova:2021nnc,Volkel:2022ewm} for how this can be done in practice. 
Note that this approach is not limited to the eigenvalue spectrum but can also map the corresponding eigenfunctions. 

Various methods can be used to compute bound states in a potential well; in the following, we use direct integration. 
Starting from an initial guess for the energy $E$, one numerically integrates the field from two distant starting points, here $\psi_{\pm}$, to a common point, at which one computes the Wronskian
\begin{align}\label{wronskian}
W(\psi_{-},\psi_{+})(x, E)  =  \psi_{-} \frac{\text{d} \psi_{+}}{\text{d}x} - \frac{\text{d} \psi_{-}}{\text{d}x} \psi_{+}\,.
\end{align}
If the initial guess is correct, the Wronskian will vanish; if not, the procedure is repeated with another guess, and it is thus a simple root-finding problem. 
Note that this numerical scheme is very fragile for QNMs due to numerical instabilities of the exponentially diverging solutions towards the horizon and spatial infinity but very robust for bound states
\footnote{We refer to the appendix of Ref.~\cite{Volkel:2022ewm} for more details on the stability and implementation.}.  

To model small perturbations in the Regge-Wheeler potential, $V(x) \rightarrow V(x) + \delta V(x)$, we adopt the commonly used P\"oschl-Teller potential
\begin{align}\label{deltaV}
\delta V(x) = \frac{\delta V_0}{\cosh(a (x-x_0))^2}\,,
\end{align}
where $\delta V_0$ scales the magnitude of the perturbation, $a$ scales its width, and $x_0$ its location. 
An important property of the P\"oschl-Teller is that it admits a finite number of bound states due to its exponential decay. 
As we will see, this is important for the asymptotic properties of the perturbed spectrum. 
While some studies~\cite{Barausse:2014tra,Cheung:2021bol,Berti:2022xfj,Cardoso:2024mrw,Ianniccari:2024ysv} motivated the presence of a bump in the Regge-Wheeler potential to qualitatively represent the presence of a shell of matter, in this work, we only use it as a tool to study spectral stability. 
Approximate spectral shifts could also be obtained with a complex continuation method, also known as eigenvalue perturbation method~\cite{Zimmerman:2014aha,Mark:2014aja,Hussain:2022ins,Li:2023ulk,Wagle:2023fwl,Ma:2024qcv,Weller:2024qvo,Li:2025fci}. 
Although it is related to Mashhoon's coordinate transformation, it requires prior knowledge of the unperturbed solutions and does not utilize bound states in the same manner as Mashhoon's method. 
We leave a detailed study of perturbative bound-state shifts for upcoming work.

\section{Results}

\begin{figure}[H]
\centering
\includegraphics[width=1.0\columnwidth]{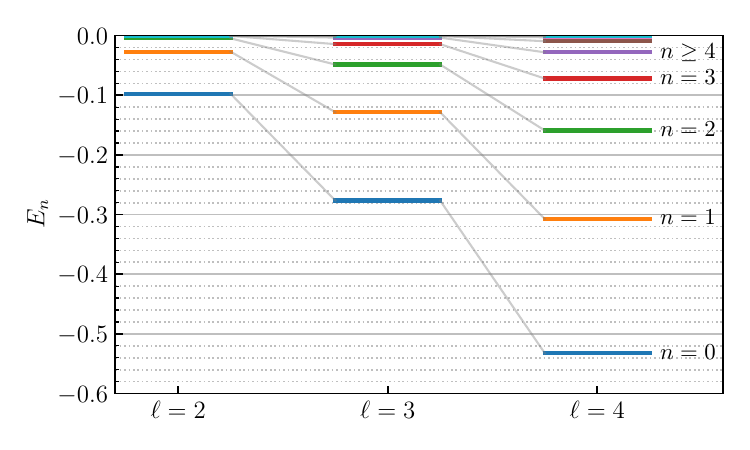}
\caption{Spectrum of bound states $E_{n}$ of the inverted Regge-Wheeler potential for $\ell \in [2,3,4]$ and $M=1$. Gray lines connect states with the same $n$ (same color) for different $\ell$. 
\label{Enall}}
\end{figure}

Before discussing our results and their implications in greater detail, we first show the spectrum of bound states of the unperturbed Regge-Wheeler potential for different $\ell$ in Fig.~\ref{Enall}. 
Increasing $\ell$ deepens the potential well, thus, the energy levels $E_n$ become more negative. 
States with increasing $n$ quickly approach 0, which has important implications for the localization of eigenfunctions and turning points, as we will see later. 

To ease the discussion, we focus on $\ell=2$ but provide similar results for $\ell=3$ and $\ell=4$ in the supplementary material. 
In Fig.~\ref{fig1}, we present eigenfunctions for the first four bound states as $\psi_n(x)$, and the corresponding energy levels $E_n$ as horizontal lines. 
Eigenfunctions are normalized such that $\int |\psi_n(x)|^2 \text{d}x = 1$. 
Additional bound states can be found in Fig.~\ref{figboundstates}. 
In the supplementary material, we provide their numerical values in Table~\ref{table1}. 

\begin{figure}[H]
\centering
\includegraphics[width=1.0\columnwidth]{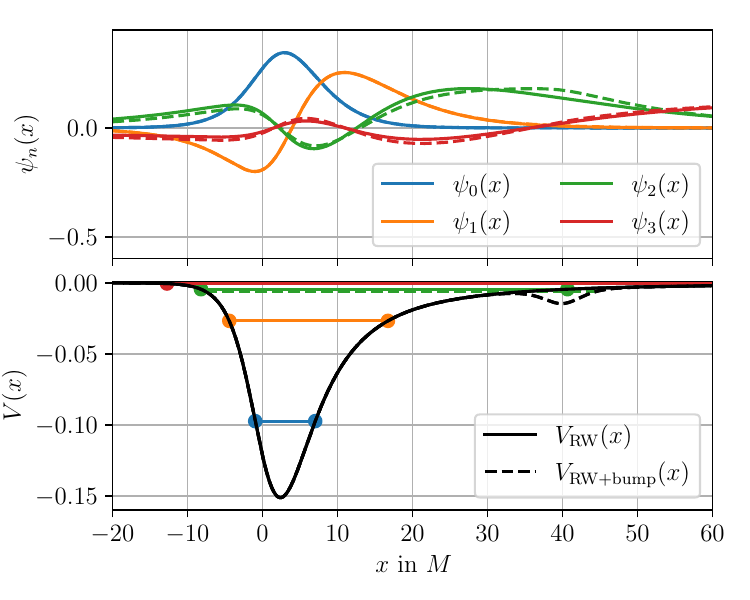}
\caption{
Inverted Regge-Wheeler potential without (black solid) and with an artificial bump (black dashed) together with the first four eigenfunctions (colored solid and dashed, respectively). The energy levels of both potentials are indicated as horizontal lines that start and end at the classical turning points of the unperturbed Regge-Wheeler potential.
\label{fig1}}
\end{figure}

\textbf{\textit{Unperturbed Regge-Wheeler potential:}} 
Let us first discuss the unperturbed Regge-Wheeler case. 
From Fig.~\ref{fig1}, we find that the $n=0$ state is well localized near the minimum of the potential; from here on called $x_\text{min}$. 
One should also expect this from the QNM picture because it is well described by the local properties of the maximum of the potential barrier, e.g., via the Schutz-Will formula~\cite{Schutz:1985km}. 

The $n>0$ states become less localized, with an apparent asymmetry towards $x > x_\text{min}$, i.e., the oscillatory region increases mostly towards large positive $x$. 
For instance, for $n=2$ the maximum of $|\psi_2(x)|^2$ is already beyond $20\,M$. 
For even larger $n$, the oscillatory region of the eigenfunctions spread out significantly further, demonstrating the wavefunction's sensitivity to the long-range properties of the potential\footnote{If one were to make the quantitative connection to quantum mechanical probability densities for a closer comparison to the hydrogen atom, note that the derivation of the master equations involves a $1/r$ rescaling which would need to be included. }.

\begin{figure}
\centering
\includegraphics[width=1.0\columnwidth]{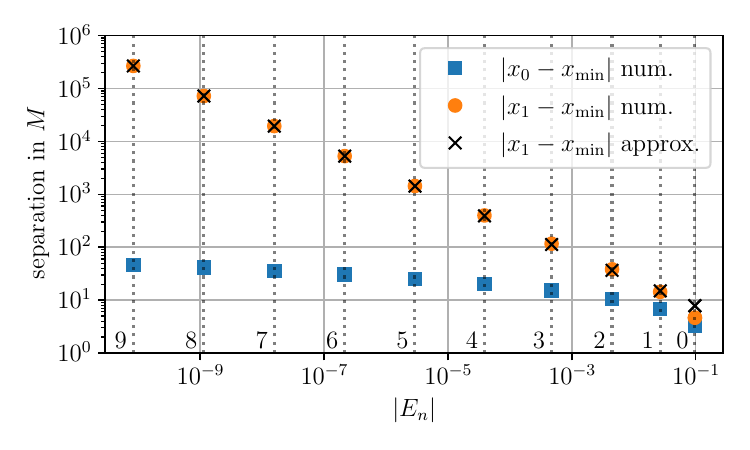}
\caption{Here we relate the Regge-Wheeler bound states $E_n$ with their associated left ($x_0(E_n)$, blue squares) and right turning points ($x_1(E_n)$, orange circles). For comparison, we also show the asymptotic approximation Eq.~\eqref{tp_approx} for the right turning point (black crosses). We indicate the value of the bound states $E_n$ (vertical black dotted) and its number. \label{figtp}}
\end{figure}

The rapid delocalization of the wavefunctions can also be explained with a naive application of the Bohr-Sommerfeld rule~\cite{bender1999advanced}, which states that the bound-state energies in a potential are determined from integrating $\sqrt{E-V(x)}$ between the two classical turning points $x_{1,2}$ defined by $E=V(x_{1,2})$. 
Because the energy levels of the inverted Regge-Wheeler potential quickly approach zero, see Table~\ref{table1}, the corresponding turning points, especially the outer one ($x_1 > x_\text{min}$), deviate rapidly from $x_\text{min}$. 
We demonstrate this scaling in Fig.~\ref{figtp}, where we also show the leading-order approximation for the outer turning point obtained by approximating the potential for large $x$, given by
\begin{align}\label{tp_approx}
x_\mathrm{1} \approx \sqrt{\ell(\ell+1)/(-E_n)}\,. 
\end{align}

From the numerical computation of the bound states $E_n$ for larger $n$, reported in Fig.~\ref{figboundstates}, it seems $E_n$ approaches zero exponentially fast. 
A closer look at the spacing between neighboring states, reported in Fig.~\ref{figspacing}, suggests that $K_n \equiv \ln(E_{n}/E_{n+1})$ approaches a constant value for large $n$. 
Since the wavefunction is mainly characterized by the outer part of the potential (see Fig.~\ref{fig1}), one could be tempted to approximate the spacing with the one of a truncated $\ell(\ell+1)/x^2$ potential. 
The asymptotic spacing of large $n$ eigenvalues in our application would be given by\footnote{It can be derived using results from Refs.~\cite{Camblong:2000ec,Coon:2002sua} for the Bessel functions. 
However, only the log separation of states is independent of the introduced truncation parameter, not the value of $E_n$ itself. 
Also, the absence of the tortoise coordinate in Refs.~\cite{Camblong:2000ec,Coon:2002sua} implies subtle differences for the number of existing bound states, i.e., the exact potential studied in Refs.~\cite{Camblong:2000ec,Coon:2002sua} has either one bound state or none; see also Ref.~\cite{Nollert:1998ys}.}
\begin{align}
\label{Enspacing}
E_{n+1} \approx E_{n} e^{-K}\,, \quad K=\frac{2 \pi}{\sqrt{\ell(\ell+1)-1/4}}\,.
\end{align}
This estimate is an excellent approximation for the Regge-Wheeler states beyond the first few $n$, as shown in Fig.~\ref{figspacing}. 
Note that the iterative application of Eq.~\eqref{Enspacing} starting from $n=0$ yields a qualitative scaling of $E_n \sim -\exp\left(- K n \right)$, but including a finite error due to limited accuracy of Eq.~\eqref{Enspacing} for small $n$. 
In Fig.~\ref{figspacing2} of the supplementary material, we show that the convergence to $K$ is even exponential. 
Note that in the eikonal (large $\ell$) limit, the spacing is simply controlled by $K\approx 2 \pi/\ell$, and thus $e^{-K}\approx 1$.  

Interestingly, the presence of the power-law tail in the time-evolution of initial data, known as the Price tail, also relies crucially on the long-range properties of the potential~\cite{Price:1972pw}. 
We leave the question of whether one can more systematically map properties of the power-law tail to the asymptotic spacing of bound states for future work. 

\textbf{\textit{Perturbed Regge-Wheeler potential:}} 
Let us now contrast these findings with the perturbed Regge-Wheeler potential, for which all relevant quantities are reported in the same figures for a direct comparison. 
As one explicit example, we set $\delta V_0=0.01$, $a=0.3$, and $x_0=40\,M$. 
A closer look at the eigenfunctions in Fig.~\ref{fig1} reveals that the $n=0$ and $n=1$ states are practically insensitive to the modification, while visible differences become present for $n=2$. 
This is expected, because $|E_1| < |\delta V_0| < |E_2|$, which we show in the top panel of Fig.~\ref{figboundstates}. 
While the overall change to $E_n$ is small, the absolute difference between the unperturbed and perturbed states, defined as $|\Delta E_n| = |E_n-E^\text{bump}_n|$, peaks for $n=2$ just below $\delta V_0$, and then decreases afterward. 
Note that the relative difference of all subsequent states, defined as $\delta |E_n| = |(E_n-E^\text{bump}_n)/E_n|$, is negligible for $n<2$, but remains at a roughly constant level given by $\delta V_0/V_0$ for $n \geq 2$. 
In Fig.~\ref{figspacing_l34} in the supplementary material, we show results for $\ell=3$ and $\ell=4$ underlining that $n=2$ does not play a special role. 

\begin{figure}
\centering
\includegraphics[width=1.0\columnwidth]{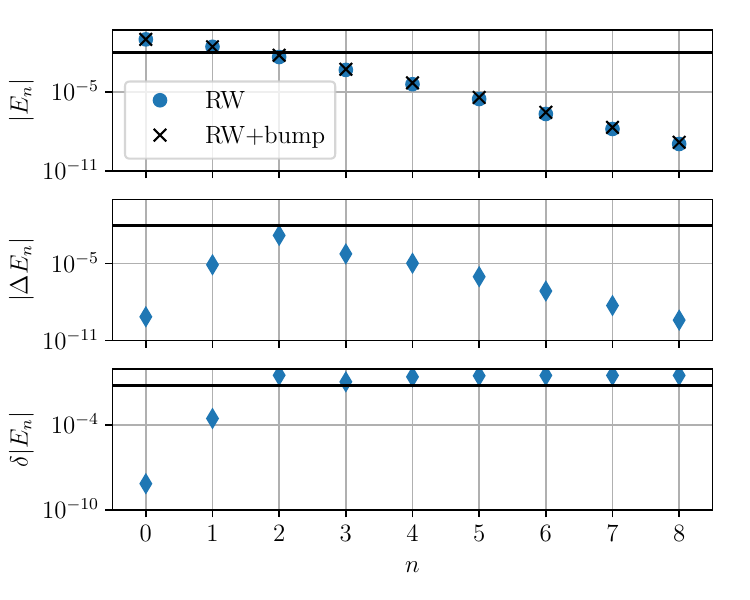}
\caption{Here we compare the bound states of the Regge-Wheeler potential with and without a bump. In the top panel, we show the actual bound states; in the middle panel, we show their absolute difference; and in the bottom panel, the relative difference. 
We indicate the minimum of the bump $\delta V_0$ in the top and middle panel (black solid line) and the ratio of $\delta V_0/V_0$ in the bottom panel (black solid line). The data are also reported in Table~\ref{table1} in the supplementary material~\ref{app1}. \label{figboundstates}}
\end{figure}

Finally, from Fig.~\ref{figspacing}, we conclude that the spacing of modes is only different for moderate $n$, but the presence of $\delta V(x)$ does not change its asymptotic properties. 
Note that this is related to the quick decay of $\delta V(x)$ and would be different if $\delta V(x)$ would change the coefficient in front of the asymptotic $1/x^2$ decay or decay even slower. 
A detailed study of the spectral stability of the bound states is left for future work. 
We refer to results reported in Refs.~\cite{Cheung:2021bol,Berti:2022xfj,Cardoso:2024mrw} for discussion on how the Schwarzschild QNMs are being impacted by various choices of the perturbing P\"oschl-Teller potential. 
Note that the trapping of waves between the two barriers in the QNM problem is qualitatively related to tunneling between potential wells in the bound-state picture. 

\begin{figure}
\centering
\includegraphics[width=1.0\columnwidth]{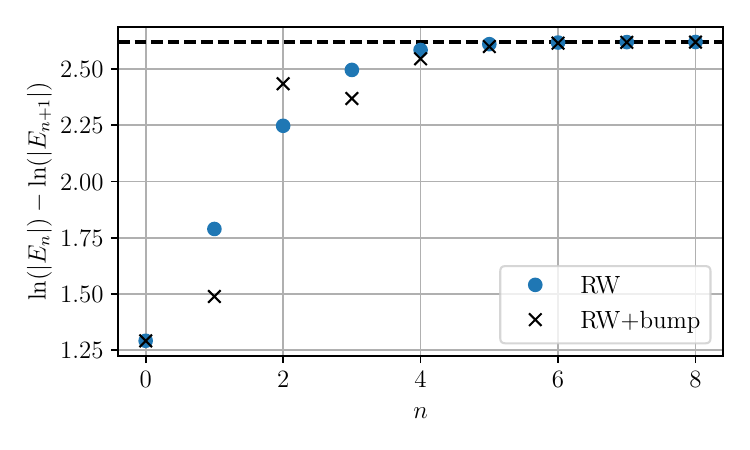}
\caption{Here we show the separation of subsequent energy levels $\ln(|E_{n}|)-\ln(|E_{n+1}|)$ for the two potentials, as well as the asymptotic spacing expected from Eq.~\eqref{Enspacing} (black dashed).
\label{figspacing}}
\end{figure}

\section{Conclusions}\label{conclusion}

In this work, we have explicitly computed bound states $E_n$ and eigenfunctions of the inverted Regge-Wheeler potential. 
To the best of our knowledge, this is the first study of this kind in the literature. 
For a compact discussion and presentation, we have mostly focused on the $\ell=2$ energy levels and corresponding eigenfunctions for the first 10 bound states, but provide additional results for $\ell\in[3,4]$ in the supplementary material. 

We find that bound states for large $n$ are approximately proportional to $E_n\sim-\exp(-K n)$, which is quite in contrast to the hydrogen atom levels of $E_n\sim-1/n^2$. 
Moreover, the oscillatory part of the eigenfunctions rapidly probes the asymptotic region of the potential well towards radial infinity. 
It is not obvious, but it is certainly interesting to point out that the outer turning point deviates even exponentially fast from the potential minimum. 
Based on Mashhoon's mapping~\cite{Mashhoon:1982im}, one would thus expect that even moderate QNM overtone numbers are extremely sensitive to long-range modifications. 
Due to the exponential scaling, one might describe them as loosely bound or even almost as a continuum. 
This is interesting in the context of Ref.~\cite{Steinhauer:2025fqp}, which discuss the connection of Schwarzschild QNMs and a continuum. 

We have also demonstrated that the bound states to QNMs correspondence can provide complementary insights to open problems in the QNM literature, in particular for spectral stability~\cite{Nollert:1996rf,Nollert:1998ys,Barausse:2014tra,Cheung:2021bol,Berti:2022xfj,Cardoso:2024mrw}. 
Because the turning points of the bound-state problem are real, it is also straightforward to relate properties of the potential well with certain bound states. 
Based on Mashhoon's identification with QNMs, it is tempting to do the same for the corresponding QNM. 
In this context, it is worth mentioning that the sensitivity of overtones to modifications of the effective potential close to the horizon has been pointed out recently in Refs.~\cite{Konoplya:2022pbc,Silva:2024ffz}. 
This finding is qualitatively understandable in the bound state picture presented here. 

Since this study is a brief report focusing on the main results, we plan to provide a more complete analysis in a separate work. 
However, the extension to $n \geq 10$  and $\ell \geq 5$ is straightforward and should provide qualitatively similar findings. 
Quantitative differences will arise due to the increasing depth of the potential well as a function of $\ell$. 
Because of isospectrality, the computation for polar perturbations should yield the same conclusions, while those for a test scalar field should be qualitatively similar. 

Finally, given the interest of the community in Mashhoon's and Maggiore's pioneering studies~\cite{Mashhoon:1982im,Maggiore:2007nq}, we hope this work motivates the exploration of bound states of black holes more systematically in the future. 
Possible directions include a quantitative connection between the power-law tail in the time evolution of initial data and the asymptotic spacing of bound states, which we briefly commented on. 
Moreover, future work could also address the role of excitation factors in the bound state picture and whether their loosely-bound nature can be related to ongoing work on extracting QNM overtones, e.g., ~\cite{Baibhav:2023clw,Nee:2023osy,Giesler:2024hcr,Thomopoulos:2025nuf}. 
Ultimately, it will be exciting to quantify how far the bound-states picture can be extended to the Kerr black hole. 

\paragraph{Acknowledgments} 
S.\,H.\,V. wants to thank Vitor Cardoso, Nicola Franchini, Guillermo Lara, Kostas D.\,Kokkotas, Raj Patil, Hector O.\,Silva and Pratik Wagle for their valuable discussions and feedback on this manuscript. 
S.\,H.\,V. is indebted to the anonymous referees for their insightful comments. 
S.\,H.\,V. acknowledges funding from the Deutsche Forschungsgemeinschaft (DFG): Project No. 386119226.

\paragraph{Data availability} 
The data supporting this study's findings are available within the Letter~\cite{volkel_2025_15577458}.

\bibliography{literature}

\clearpage
\appendix

\section{Supplementary material}\label{app1}

\subsection{Convergence of $K$}
\begin{figure}[H]
\centering
\includegraphics[width=1.0\columnwidth]{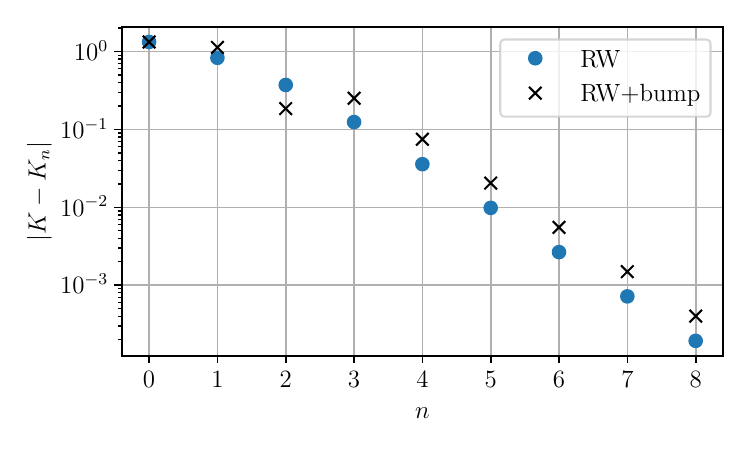}
\caption{Here, we show the exponential convergence of $|K-K_n|$ for the Regge-Wheeler potential with and without a bump.
\label{figspacing2}}
\end{figure}

In Fig.~\ref{figspacing2}, we show the convergence of $K_n$ to $K$ from Eq.~\eqref{Enspacing} for the two potentials. 
Note that for large $n\geq3$ the slope is very similar for both potentials. 

\subsection{Additional results for $\ell=3$ and $\ell=4$}\label{app1_2}
\begin{figure}[H]
\centering
\includegraphics[width=1.0\columnwidth]{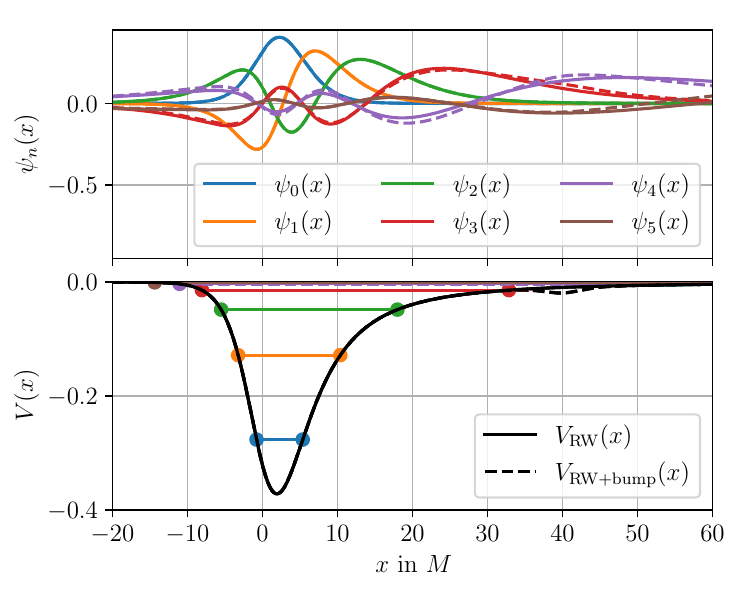}
\includegraphics[width=1.0\columnwidth]{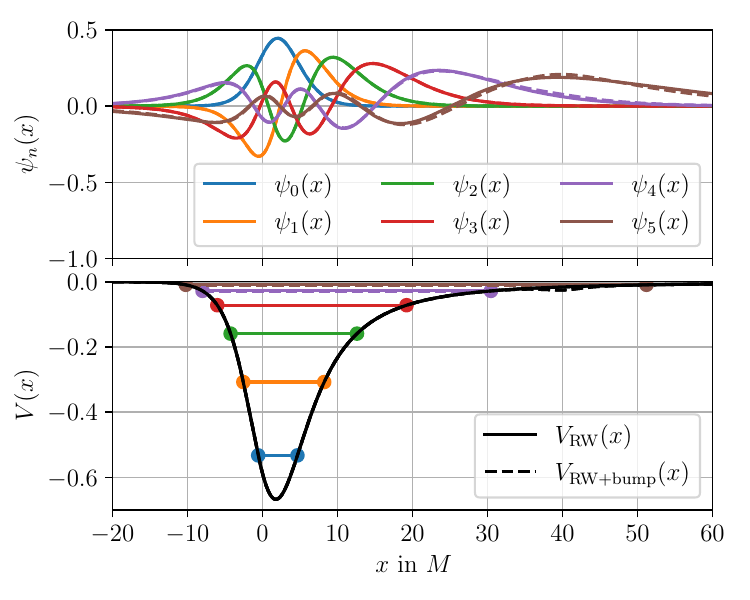}
\caption{
Inverted Regge-Wheeler potential for $\ell=3$ (top panel) and $\ell=4$ (bottom panel) without (black solid) and with an artificial bump at $x_0=40\,M $(black dashed) together with the first six eigenfunctions (colored solid and dashed, respectively). The energy levels of both potentials are indicated as horizontal lines that start and end at the classical turning points of the unperturbed Regge-Wheeler potential.
\label{fig1_l34}}
\end{figure}

\begin{figure}[H]
\centering
\includegraphics[width=1.0\columnwidth]{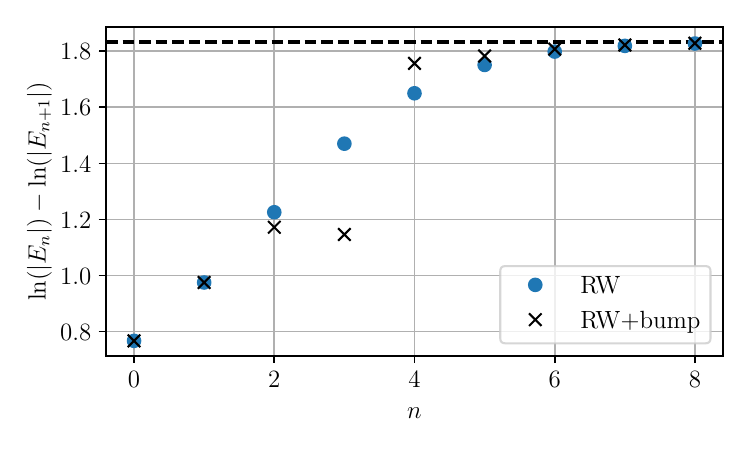}
\includegraphics[width=1.0\columnwidth]{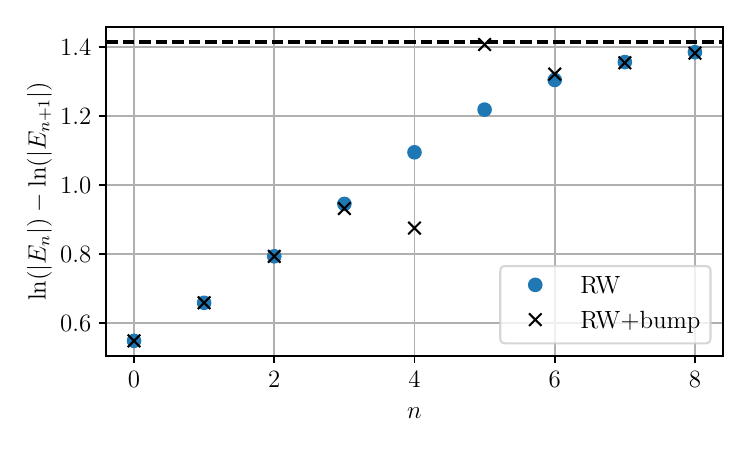}
\caption{Here we show the separation of subsequent energy levels $\ln(|E_{n}|)-\ln(|E_{n+1}|)$ for the two potentials, as well as the asymptotic spacing expected from Eq.~\eqref{Enspacing} (black dashed) for $\ell=3$ (top panel) and $\ell=4$ (bottom panel).
\label{figspacing_l34}}
\end{figure}

In Fig.~\ref{fig1_l34}, we show the first 6 eigenfunctions of the inverted Regge-Wheeler potential (without and with perturbation) for $\ell=3$ and $\ell=4$. 
Note that the eigenfunctions become more localized for larger $\ell$ (for the same $n$). 
In Fig.~\ref{figspacing_l34}, we show the corresponding spacing of bound states along with the asymptotic prediction Eq.~\eqref{Enspacing}. 
Note that it takes larger $n$ to approach the asymptotic prediction when increasing $\ell$ and that the shifts due to the perturbation in the potential also appear for larger $n$. 
\vfill\break

\subsection{Tabulated bound states}

In Table~\ref{table1}, we report the numerical results for the bound states of the Regge-Wheeler potential and its perturbation ($\delta V_0=0.01$, $a=0.3$, and $x_0=40\,M$) considered in the main text for $\ell \in[2,3,4]$. 

\vfill\break

\begin{table}[H]
\[
\begin{array}{cccc}
\hline
\hline
\multicolumn{4}{c}{\ell=2} \\
\hline
 n & |E_n|  & |E^\text{bump}_n| & |\delta E_n|\\
\hline
0 & 9.741e-02 & 9.741e-02 & 7.321e-09 \\ 
1 & 2.677e-02 & 2.678e-02 & 3.034e-04 \\ 
2 & 4.474e-03 & 6.042e-03 & 3.503e-01 \\ 
3 & 4.728e-04 & 5.295e-04 & 1.199e-01 \\ 
4 & 3.897e-05 & 4.957e-05 & 2.720e-01 \\ 
5 & 2.940e-06 & 3.889e-06 & 3.228e-01 \\ 
6 & 2.161e-07 & 2.889e-07 & 3.370e-01 \\ 
7 & 1.577e-08 & 2.114e-08 & 3.409e-01 \\ 
8 & 1.149e-09 & 1.541e-09 & 3.419e-01 \\ 
9 & 8.361e-11 & 1.122e-10 & 3.422e-01 \\ 
\hline
\end{array}
\]

\[
\begin{array}{cccc}
\hline
\hline
\multicolumn{4}{c}{\ell=3} \\
\hline
 n & |E_n|  & |E^\text{bump}_n| & |\delta E_n|\\
\hline
0 & 2.764e-01 & 2.764e-01 & 9.795e-11 \\ 
1 & 1.283e-01 & 1.283e-01 & 1.389e-08 \\ 
2 & 4.840e-02 & 4.840e-02 & 5.855e-05 \\ 
3 & 1.421e-02 & 1.499e-02 & 5.507e-02 \\ 
4 & 3.266e-03 & 4.762e-03 & 4.583e-01 \\ 
5 & 6.273e-04 & 8.224e-04 & 3.111e-01 \\ 
6 & 1.089e-04 & 1.383e-04 & 2.701e-01 \\ 
7 & 1.802e-05 & 2.269e-05 & 2.587e-01 \\ 
8 & 2.923e-06 & 3.669e-06 & 2.550e-01 \\ 
9 & 4.702e-07 & 5.894e-07 & 2.536e-01 \\ 
\hline
\end{array}
\]

\[
\begin{array}{cccc}
\hline
\hline
\multicolumn{4}{c}{\ell=4} \\
\hline
 n & |E_n|  & |E^\text{bump}_n| & |\delta E_n|\\
\hline
0 & 5.325e-01 & 5.325e-01 & 2.477e-11 \\ 
1 & 3.075e-01 & 3.075e-01 & 3.790e-10 \\ 
2 & 1.591e-01 & 1.591e-01 & 2.586e-08 \\ 
3 & 7.193e-02 & 7.193e-02 & 2.427e-05 \\ 
4 & 2.795e-02 & 2.833e-02 & 1.365e-02 \\ 
5 & 9.354e-03 & 1.180e-02 & 2.619e-01 \\ 
6 & 2.767e-03 & 2.892e-03 & 4.530e-02 \\ 
7 & 7.512e-04 & 7.721e-04 & 2.785e-02 \\ 
8 & 1.937e-04 & 1.995e-04 & 3.009e-02 \\ 
9 & 4.852e-05 & 5.012e-05 & 3.295e-02 \\ 
\hline
\end{array}
\]
\caption{
Here, we report the first ten bound states $E_n$ for $\ell\in[2,3,4]$ of the Regge-Wheeler potential, the ones of the perturbed potential $E^\text{bump}_n$, along with the relative difference $\delta E_n$ between the two spectra ($M=1$). 
}
\label{table1}
\end{table}

\end{document}